\documentclass[aps,prb,preprint,superscriptaddress,nofootinbib]{revtex4-1}
\usepackage{color}
\usepackage{graphicx}
\usepackage{lineno}
\usepackage[normalem]{ulem}
\usepackage[colorinlistoftodos]{todonotes}
\newcommand{\BFAP}{BaFe$_2$(As$_{1-x}$P$_x$)$_2$~}
\newcommand{\BFA}{BaFe$_2$As$_2$~}

\begin{document}
\title{Reciprocity between local moments and collective magnetic excitations in the phase diagram of \BFAP}

\author{Jonathan Pelliciari\footnote{pelliciari@bnl.gov}}
\affiliation{Swiss Light Source, Photon Science Division, Paul Scherrer Institut, CH-5232 Villigen PSI, Switzerland}
\affiliation{Department of Physics, Massachusetts Institute of Technology, Cambridge, Massachusetts 02139, USA}
\affiliation{NSLS-II, Brookhaven National Laboratory, Upton, NY, 11973, USA}
\author{Kenji Ishii}
\affiliation{Synchrotron Radiation Research Center, National Institutes for Quantum and Radiological Science and Technology, Sayo, Hyogo 679-5148, Japan}
\author{Yaobo Huang}
\affiliation{Swiss Light Source, Photon Science Division, Paul Scherrer Institut, CH-5232 Villigen PSI, Switzerland}
\affiliation{Beijing National Lab for Condensed Matter Physics, Institute of Physics, Chinese Academy of Sciences P. O. Box 603, Beijing 100190, China}
\author{Marcus Dantz}
\affiliation{Swiss Light Source, Photon Science Division, Paul Scherrer Institut, CH-5232 Villigen PSI, Switzerland}
\author{Xingye Lu}
\affiliation{Swiss Light Source, Photon Science Division, Paul Scherrer Institut, CH-5232 Villigen PSI, Switzerland}
\author{Paul Olalde Velasco\footnote{Current address: Diamond Light Source, Harwell Science and Innovation Campus, Didcot, Oxfordshire, OX11 0DE, United Kingdom}}
\affiliation{Swiss Light Source, Photon Science Division, Paul Scherrer Institut, CH-5232 Villigen PSI, Switzerland}
\author{Vladimir N. Strocov}
\affiliation{Swiss Light Source, Photon Science Division, Paul Scherrer Institut, CH-5232 Villigen PSI, Switzerland}
\author{Shigeru Kasahara}
\affiliation{Department of Physics, Kyoto University, Sakyo-ku, Kyoto 606-8502, Japan}
\author{Lingyi Xing}
\affiliation{Beijing National Lab for Condensed Matter Physics, Institute of Physics, Chinese Academy of Sciences, Beijing 100190, China}
\author{Xiancheng Wang}
\affiliation{Beijing National Lab for Condensed Matter Physics, Institute of Physics, Chinese Academy of Sciences, Beijing 100190, China}
\author{Changqing Jin}
\affiliation{Beijing National Lab for Condensed Matter Physics, Institute of Physics, Chinese Academy of Sciences, Beijing 100190, China}
\affiliation{Collaborative Innovation Center for Quantum Matters, Beijing, China}
\author{Yuji Matsuda}
\affiliation{Department of Physics, Kyoto University, Sakyo-ku, Kyoto 606-8502, Japan}
\author{Takasada Shibauchi}
\affiliation{Department of Advanced Materials Science, University of Tokyo, Kashiwa, Chiba 277-8561, Japan}
\author{Tanmoy Das}
\affiliation{Department of Physics, Indian Institute of Science, Bangalore-560012, India}
\author{Thorsten Schmitt\footnote{thorsten.schmitt@psi.ch}}
\affiliation{Swiss Light Source, Photon Science Division, Paul Scherrer Institut, CH-5232 Villigen PSI, Switzerland}

\begin{abstract}
Unconventional superconductivity arises at the border between the strong coupling regime with local magnetic moments and the weak coupling regime with itinerant electrons, and stems from the physics of criticality that dissects the two. Unveiling the nature of the quasiparticles close to quantum criticality is fundamental to understand the phase diagram of quantum materials. Here, using resonant inelastic x-ray scattering (RIXS) and Fe-K$_\beta$ emission spectroscopy (XES), we visualize the coexistence and evolution of local magnetic moments and collective spin excitations across the superconducting dome in isovalently-doped \BFAP (0.00$\leq$x$\leq0.$52). Collective magnetic excitations resolved by RIXS are gradually hardened, whereas XES reveals a strong suppression of the local magnetic moment upon doping. This relationship is captured by an intermediate coupling theory, explicitly accounting for the partially localized and itinerant nature of the electrons in Fe pnictides. Finally, our work identifies a local-itinerant spin fluctuations channel through which the local moments transfer spin excitations to the particle-hole (paramagnons) continuum across the superconducting dome.
\end{abstract}
\maketitle

\section*{Introduction}
\begin{figure}
\includegraphics[width=\textwidth,trim={0 540 0 0},clip]{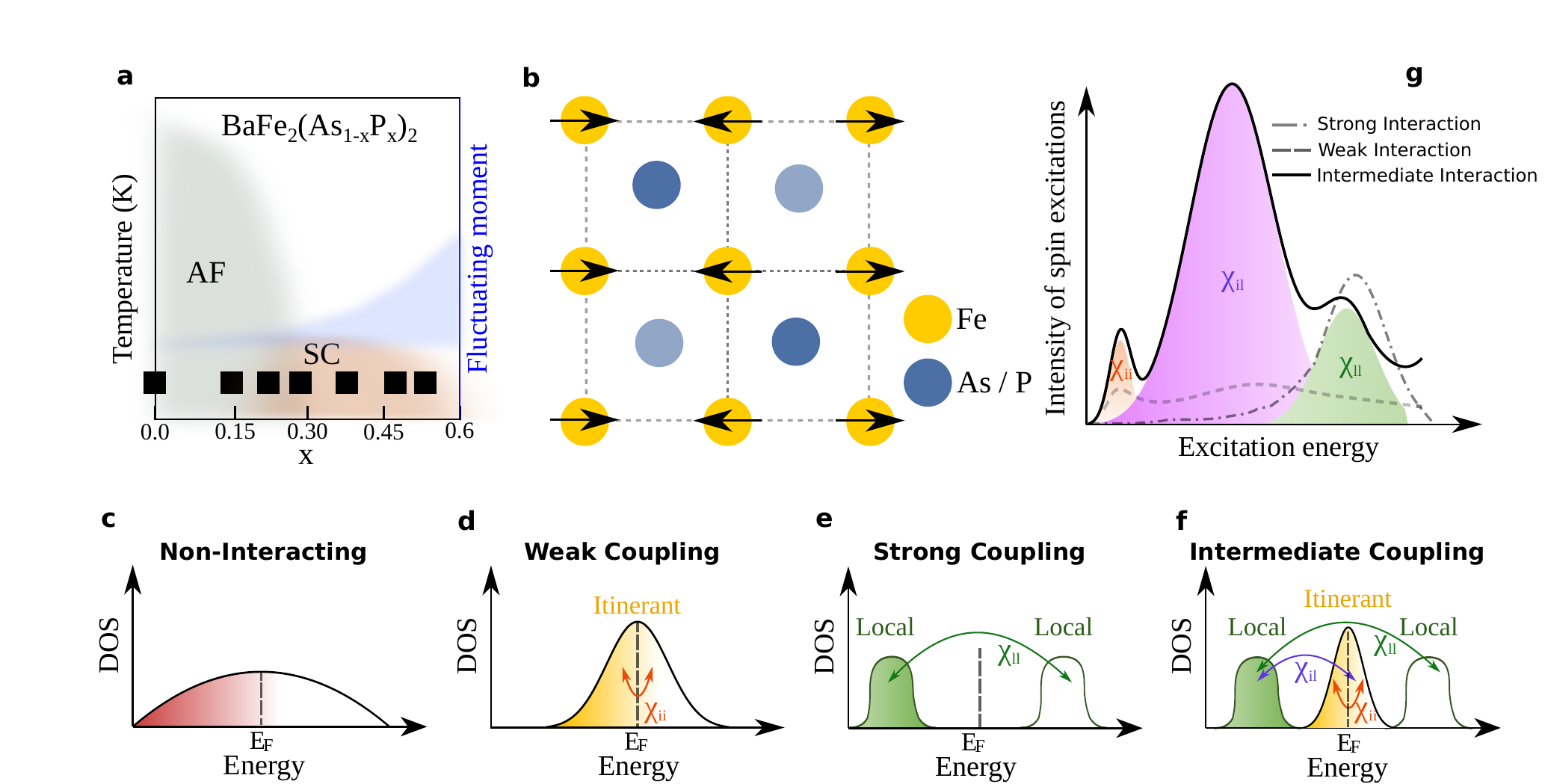}
\caption{\label{fig:fig1} \textbf{Phase diagram, structure, and exictations of \BFAP}. \textbf{a} Phase diagram of \BFAP. The black squares represent the doping levels and temperature measured in our work. As blue line, we schematically depict the expected behaviour of the fluctuating moment. \textbf{b} Schematic real space magnetic structure of \BFA. \textbf{c-f} Density of states (DOS) as a function of quasiparticle energy. Non-interacting \textbf{c}, weak-coupling Random Phase Approximation (RPA) approximation \textbf{d}, strong coupling \textbf{e}, and intermediate coupling Momentum Resolved Density Fluctation theory (MRDF) \textbf{f}. \textbf{g} Excitations spectrum of  the spin excitations arising from weak, strong, and intermediate coupling.}
\end{figure}

It is now well established that unconventional superconductivity (SC) originates from a unique and augmented manifestation of the electronic correlation properties that arises as the system is driven towards a quantum critical region via various tuning parameters such as doping and pressure \cite{johnston_puzzle_2010,stewart_superconductivity_2011,chubukov_pairing_2012,scalapino_common_2012,mannella_magnetic_2014,inosov_spin_nodate,tranquada_superconductivity_2014,dai_antiferromagnetic_2015,dai_magnetism_2012,lumsden_magnetism_2010,shibauchi_quantum_2014}. On one side of the superconducting dome, the correlation strength is strongly enhanced, leading to diverse quantum many-body effects such as the Mott insulating state, non-Fermi-liquid behaviour, and magnetic orders (see Fig.~\ref{fig:fig1}\textbf{a,b}) \cite{lee_doping_2006,chubukov_pairing_2012,scalapino_common_2012,yin_kinetic_2011,de_medici_orbital-selective_2009}. On the other side of the dome, the correlation strength is often substantially suppressed, and the low-energy physics can be described by a more conventional Fermi-liquid theory \cite{shibauchi_quantum_2014}. The transition region between these two limits of correlation holds a quantum critical point barely understood. Generally, SC is optimized in this intermediate region where the cooperation of a strongly enhanced non-Fermi-liquid behaviour and the presence of a quantum criticality leads to unconventional and not understood physics \cite{shibauchi_quantum_2014}. This generic phase diagram suggests that intertwined electronic and magnetic instabilities, arising from the intermediate correlation strength close to this critical region, induce strong Cooper pairing \cite{chubukov_pairing_2012,scalapino_common_2012,chubukov_itinerant_2015}.

The physics which determines the evolution of the spin excitations is crucially dependent on the interaction strength of the electrons compared to their bandwidth. We explain this phenomenon schematically in Fig.~\ref{fig:fig1}\textbf{c-g}. The non-interacting density of states (DOS) of electrons (Fig.~\ref{fig:fig1}\textbf{c}) is renormalized by the interaction in the two extreme limits as: (i) the bands become sharper in the weak coupling quasiparticle picture (Fig.~\ref{fig:fig1}\textbf{d}) and (ii) split into two Mott-like bands characterized by local moments in the strong coupling limit (Fig.~\ref{fig:fig1}\textbf{e}). (i) In the weak coupling limit, the low-energy spin excitations ($\chi_{ii}$, see Fig.~\ref{fig:fig1}\textbf{d}) are very fragile and mix with the particle-hole continuum, failing to form a localized moment. (ii) In the strong coupling limit, the spin excitations across the Mott-bands ($\chi_{ll}$, see Fig.~\ref{fig:fig1}\textbf{e}) feature a gapped behaviour on the order of the onsite energy U without the particle-hole continuum\cite{yin_kinetic_2011,de_medici_orbital-selective_2009}. However, as the correlation strength is tuned to the intermediate coupling region (Fig.~\ref{fig:fig1}\textbf{f}), the correlated DOS co-hosts the quasiparticle DOS at low-energy and Mott states at high-energy. In this way a new spin-excitation channel ($\chi_{il}$ see Fig.~\ref{fig:fig1}\textbf{g}) appears, through which the local moment can now decay to the particle-hole (paramagnon) channels across the magnetic quantum critical point (QCP). Fe pnictides (FePns) can be placed in this intermediate region where the interplay of local and itinerant electronic states leads to high temperature SC \cite{shibauchi_quantum_2014,chubukov_itinerant_2015,johnston_puzzle_2010,stewart_superconductivity_2011}. An important open question in this context is how local magnetic moments and collective spin excitations are evolving across the superconducting dome in FePns.

FePns have a layered structure that reduces dimensionality and the parent compounds exhibit a spin-density wave with collinear antiferromagnetic (AF) order (see Fig.~\ref{fig:fig1}\textbf{b}), which gives way to SC via doping as outlined in Fig.~\ref{fig:fig1}\textbf{a} \cite{johnston_puzzle_2010,stewart_superconductivity_2011,chubukov_pairing_2012,scalapino_common_2012,mannella_magnetic_2014,inosov_spin_nodate,tranquada_superconductivity_2014,dai_antiferromagnetic_2015,dai_magnetism_2012,lumsden_magnetism_2010,shibauchi_quantum_2014}. The origin of the magnetism is poised between being itinerant, as in Cr, and localized, as in cuprates or heavy fermion materials \cite{johnston_puzzle_2010,stewart_superconductivity_2011,mannella_magnetic_2014,inosov_spin_nodate,tranquada_superconductivity_2014,dai_magnetism_2012,dai_antiferromagnetic_2015,lumsden_magnetism_2010,shibauchi_quantum_2014,yang_evidence_2009} producing an uncommon behaviour which has important consequences for the properties of FePns. Theoretical models proposed that the pairing interaction leading to the superconducting phase is provided by residual AF fluctuations persisting upon doping \cite{scalapino_common_2012,chubukov_itinerant_2015,chubukov_pairing_2012}. Yet, owing to the contribution of the AF fluctuations from local and itinerant states, a complex interplay between them arises, which is believed to play a crucial role in shaping the superconducting dome. Thus, the experimental study of magnetism across the phase diagram of high temperature superconductors is of vital importance to provide a solid basis for testing these theories. 

Inelastic neutron scattering (INS) is the traditional technique of choice to study magnetism, being able to detect magnetic fluctuations in the full Brillouin zone (BZ) \cite{dai_magnetism_2012,dai_antiferromagnetic_2015}. Recently, Resonant Inelastic X-ray Scattering (RIXS) has emerged as a complementary technique to INS by detecting spin excitations in FePns close to the $\Gamma$ point as summarized in Refs.\cite{zhou_persistent_2013,pelliciari_presence_2016,pelliciari_local_2017,pelliciari_intralayer_2016,rahn_paramagnon_2019,garcia_anisotropic_2019}. The detection of spin excitations is enabled in RIXS thanks to the spin-orbit coupling of the intermediate state mixing the quantum numbers L and S thereby activating a channel for the detection of magnetic excitations \cite{ament_resonant_2011,jia_using_2016}.
An important consideration when comparing RIXS with INS is the portion of BZ probed by the two techniques, close to the $\Gamma$ point in the case of RIXS and at the AF wave-vector for the case of INS. Depending on the case this two positions in momentum space can be equivalent or not. Moreover, it is hard to estimate and compare the absolute weight of magnetic excitations in these two regions of BZ, but it is generally accepted that the intensity at the AF wave-vector is higher than close to the $\Gamma$ point.

In this article, we use RIXS to systematically unveil the persistence and gradual hardening of the spin excitations in isovalently-doped \BFAP across the phase diagram (see Fig.~\ref{fig:fig1}\textbf{a} for a graphical description of the doping levels). Upon doping and without nominally injecting charge carriers, the spectral weight of the spin excitations increases. Our RIXS measurements are complemented with theory placing the Fe-based superconductors in the intermediate region of correlations and well describing the evolution of the spin excitations with the spin susceptibility ($\chi_s$). Additionally, an investigation of the local fluctuating magnetic moment unravels a decrease of the local moment ($\mu_{bare}$) as a function of doping. This apparent dichotomy implies that there is a transfer of magnetic spectral weight from localized to itinerant as imposed by sum rules relations. We argue that the balance between localized and itinerant states is the key to achieve SC and plays an important role for the physics of criticality in \BFAP.

\section*{Results}
\subsection*{Resonant Inelastic X-Ray Scattering (RIXS)}
\begin{figure}
\includegraphics[width=\textwidth,trim={0 440 0 0},clip]{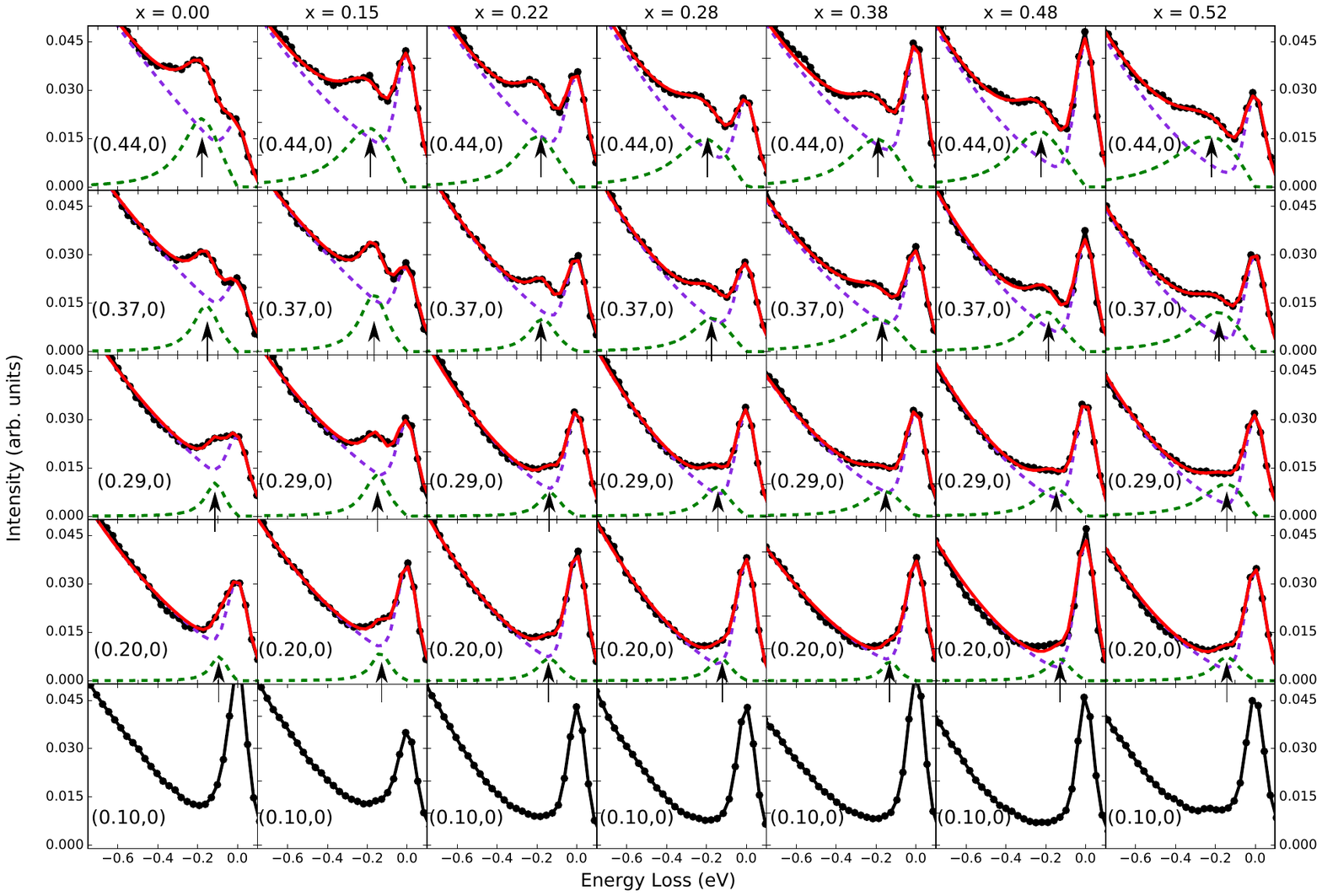}
\caption{\label{fig:fig2} \textbf{Resonant Inelastic X-ray Scattering (RIXS) spectra at low energy loss for \BFAP along (0, 0)$\rightarrow$(H, 0)}. Momentum dependence of RIXS spectra along (0, 0)$\rightarrow$(0.44, 0) for $x=$0.00, 0.15,0.22, 0.28, 0.38, 0.48, and 0.52. The incoming photons are $\pi$ polarised and the energy is tuned to the maximum of the Fe L$_3$ absorption edge. Experimental data are shown as black dots, background, and elastic peak as purple dashed line and magnetic peaks as green dashed line. The sum of background, elastic and magnetic peaks is depicted as red solid line. At low q$_{//}$ a fitting is unreliable, so no fitting was attempted.}
\end{figure}

\begin{figure}
\centering
\includegraphics[scale=0.8,trim={0 555 260 0},clip]{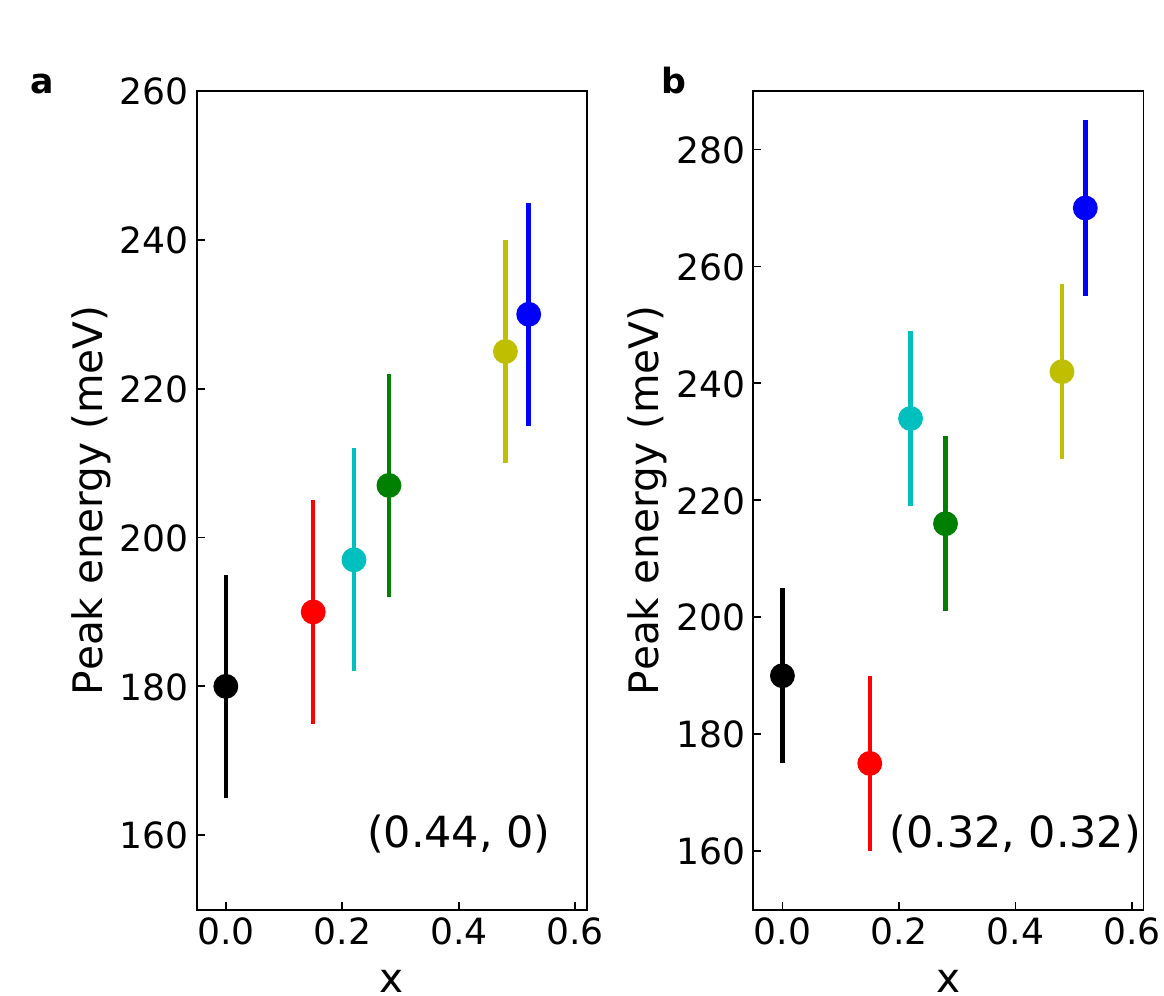}
\caption{\label{fig:fig5} \textbf{Summary of spin excitations.} \textbf{a} Dots with error bars: Maximum of the spin excitations' peak detected by Resonant Inelastic X-ray Scattering (RIXS) at (0.44,0).  \textbf{b} Dots with error bars: Maximum of the spin excitations' peak detected by RIXS at (0.32,0.32). The error bars are defined by the uncertainty of locating the zero energy position in the raw RIXS spectra. We took conservatively 30 meV (1 pixel of our  detector) which is much larger than the error resulting from the fitting analysis. }
\end{figure}

\begin{figure}
\includegraphics[width=\textwidth,trim={0 420 0 0},clip]{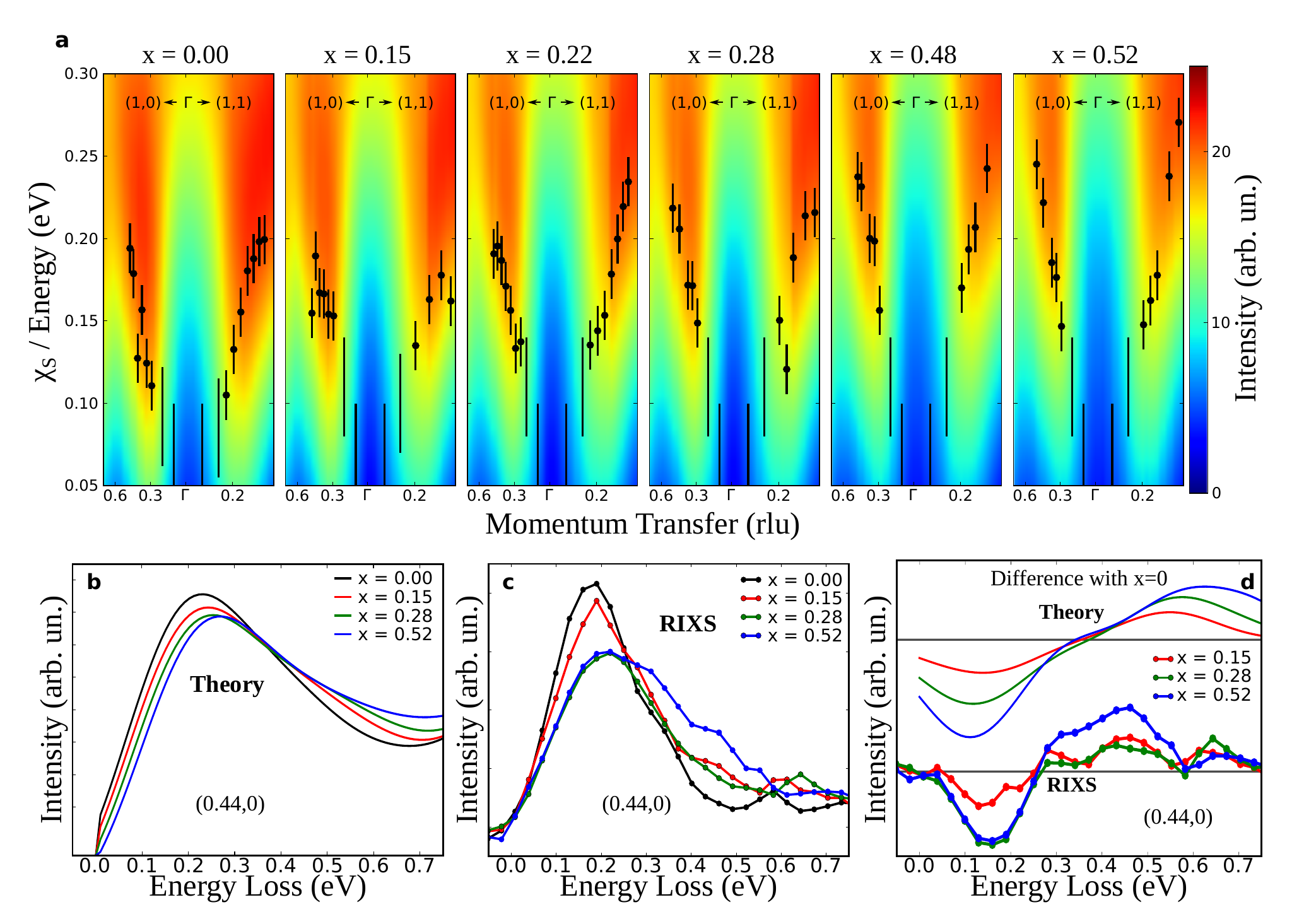}
\caption{\label{fig:fig3} \textbf{Evolution of spin excitations.} \textbf{a} Color map: Spin susceptibility ($\chi_s$) calculated by means of Momentum Resolved Density Fluctuations theory (MRDF) self-energy corrected Density Functional Theory (DFT) calculations. Black dots with error bars: Maximum of the spin excitations peak detected by Resonant Inelastic X-ray Scattering (RIXS). The error bars are defined by the uncertainty of locating the zero energy position in the raw RIXS spectra. We took conservatively 30 meV (1 pixel of our  detector) which is much larger than the error resulting from the fitting analysis. \textbf{b} Evolution of the excitations spectrum with doping in the MRDF framework at (0.44,0). \textbf{c} Peak of spin excitations from RIXS for $x=$0.00, 0.15, 0.28, and 0.52 at (0.44,0). \textbf {d} Difference of the spin excitations between doped and parent compound of RIXS spectra and MRDF calculations. }
\end{figure}

In Fig.~\ref{fig:fig2}, we show a selection of raw RIXS spectra and fitting of the elastic line, fluorescence background, and spin excitations for all the doping levels. The RIXS spectra display a low energy mode ascribed to spin excitations dispersing as a function of in plane momentum transfer (q$_{//}$) \cite{zhou_persistent_2013,pelliciari_intralayer_2016,pelliciari_presence_2016,pelliciari_local_2017,rahn_paramagnon_2019}. The bandwidth, better appreciated at high momentum transfer (top row in Fig.~\ref{fig:fig2}), gradually increases with doping along both (H,0) and (H,H) as summarized for the highest q points in Fig.~\ref{fig:fig5}.
The doping increases the width (damping) of the spin excitations leading to broader dispersing modes at high doping levels. This is expected as the hardening of the spin excitations coupled to electron-hole pair excitations naturally leads to further broadening of the spin excitations \cite{tohyama_enhanced_2015}. 
Our main results are presented in Fig.~\ref{fig:fig3}\textbf{a}, where we report the doping dependence of the dispersion of the spin excitations extracted from RIXS experiments as dots with error bars, and overlaid on the calculated dynamical spin susceptibility ($\chi_s$). All the doping levels investigated display dispersive spin excitations as shown in Fig.~\ref{fig:fig2} and~\ref{fig:fig3}\textbf{a} and in the Supplementary Figure 6. Upon doping the spin excitations, observed in our RIXS experiments, harden in energy by about 45 meV between the parent (x=0.0) compound and the most doped compound (x=0.52) at (0.44,0). The hardening for the same doping levels at (0.32,0.32) is on the order of 65 meV as summarized in Fig.~\ref{fig:fig5}.

The hardening of the spin excitations has been observed by INS for BaFe$_2$(As$_{0.7}$P$_{0.3}$)$_2$ \cite{hu_spin_2016}. At the zone boundary INS observed a hardening of the spin excitations from 180 meV (x=0) to 220 meV (x=0.3) very similar to what is detected in our RIXS data (180 meV to about 205 meV at [0.44, 0]). The small difference between INS and RIXS emerges from slightly different positions in reciprocal space ([0.5,0] vs. [0.44,0]) and doping levels (x=0.3 vs. x=0.28). The comparison to these INS measurements establishes that RIXS probes spin excitations in FePns similarly to INS, even if mixing with other channels such as charge and orbital needs to be considered \cite{nomura_resonant_2016,kaneshita_spin_2011}. The broadening effects as a function of doping detected by our RIXS experiments have been observed by INS as well, where the local susceptibility clearly shows a synergy of hardening and broadening for x=0.3.

The persistence of spin excitations along the superconducting dome was previously observed in both electron- and hole-doped FePns but with different effects on the energy of these modes. In hole-doped Ba$_{1-x}$K$_x$Fe$_2$As$_2$, the spin excitations soften upon doping due to the increase of the electronic correlations as demonstrated by theoretical calculations and accurate measurements of the Sommerfeld constant \cite{wang_doping_2013,eilers_strain-driven_2016,charnukha_intrinsic_2018} . In electron-doped FePns a different behaviour has been detected with the spin excitations being unaffected by doping in their bandwidth but with a decrease of spectral weight \cite{luo_electron_2013,wang_doping_2013}. In our present work, we uncover that the energy of the spin excitations increases upon isovalent doping and do not directly correlate with the critical temperature (T$_c$). This indicates that T$_c$ is likely connected to other microscopic details aside the effective exchange constant. Thus, the case of isovalent \BFAP reveals an unprecedented behaviour, in stark contrast to hole- and electron-doped \BFA, adding new features to the large diversity in the doping effects and magnetism in FePns.

The persistence of spin excitations in unconventional superconductors outside the antiferromagnetic parent compound is also a hallmark of the cuprates \cite{dean_persistence_2013,le_tacon_intense_2011,lee_asymmetry_2014,ishii_high-energy_2014,dellea_spin_2017}, where paramagnons have been observed across all the phase diagram. In cuprates, the energy evolution of the spin excitations upon doping is different than the case of FePns: in hole-doped cuprates there is a more or less constancy of the bandwidth along the antinodal direction whereas electron-doped cuprates display hardening of the magnetic excitations \cite{ishii_high-energy_2014,lee_asymmetry_2014,dellea_spin_2017}.

\subsection*{Momentum Resolved Density Fluctuations theory}
One approach able to describe the regime of intermediate coupling is Momentum Resolved Density Fluctuations (MRDF) theory. MRDF dissects the electronic spectrum into strongly renormalized, incoherent `local' states, and low-energy, itinerant Bloch states as displayed in Fig.~\ref{fig:fig1}\textbf{f}. The back-reaction of the quantum fluctuations to the electronic states leads to momentum, energy, and orbital dependent renormalization as well as lifetime broadening. Within the quantum field theory, these effects are captured by the real and imaginary part of the self-energy correction \cite{das_intermediate_2014}. MRDF theory self-consistently computes the dynamical correlation functions and the corresponding self-energy in the full momentum and energy space \cite{das_intermediate_2014, chubukov_itinerant_2015}. This theory captures the momentum and energy dependent evolution of the localized states as well as the dispersive quasiparticle states, resulting in an anisotropic pairing symmetry, compatible with what is observed experimentally \cite{chubukov_pairing_2012,scalapino_common_2012}. 

To describe the spin excitations probed by RIXS, the dynamical susceptibility ($\chi$) has been calculated and decomposed into the spin ($\chi_s$) and charge ($\chi_c$) channels with both the Random Phase Approximation (RPA) and MRDF method. The intensity of $\chi_s$ dominates over $\chi_c$ in agreement with a previous work \cite{pelliciari_intralayer_2016}. Density Functional Theory (DFT)-RPA calculations fail to distribute the spectral weight between the local and itinerant states and overestimate the peak energy of the spin excitations compared to our experimental data (see  Supplementary Figs. 10 and 11 and Supplementary Notes 1 and 2). The reason for this is that DFT-RPA underestimates the electronic correlations strength placing the spin excitations at too high energy. Moreover, these RPA calculations cannot account for the hardening of the spin excitations dispersion observed upon doping.
In Fig.~\ref{fig:fig1}\textbf{g}, we show an idealized scheme of the excitations pattern achieved in the intermediate coupling region comprising local-local ($\chi_{ll}$), local-itinerant ($\chi_{il}$), and itinerant-itinerant excitations ($\chi_{ii}$). The low energy $\chi_{ii}$ excitations have an energy in the order of the spin gap ($\approx$1-10 meV) and are not observable by RIXS due to current limitations of the energy resolution but are observed in INS \cite{dai_antiferromagnetic_2015,dai_magnetism_2012}. The high energy $\chi_{ll}$ appears at energies of 900-1500 meV and are not detectable by RIXS due to the intense fluorescence background. The $\chi_{il}$ excitations are the (para)-magnon excitations and can be qualitatively compared with our RIXS measurements as a function of doping. 

In Fig.~\ref{fig:fig3}\textbf{b,c}, we depict the doping dependence of $\chi_s$ directly extracted from the MRDF calculations as well as the RIXS spectra at (0.44,0). To better visualize the renormalization in energy of the magnetic excitations we also take the difference between selected doping levels and the parent compound and present the results in Fig.~\ref{fig:fig3}\textbf{d} for both theory and experiments. The agreement between theory and experiment is remarkable individuating the MRDF and more in general the intermediate coupling approach as appropriate to describe the magnetism and electronic structure of FePns.
In previous works DFT-RPA has been employed to successfully describe the spin excitations in overdoped cuprates \cite{monney_resonant_2016, guarise_anisotropic_2014,dean_itinerant_2014}. The failure of DFT-RPA to account for the spin excitations in \BFAP implies that the FePns are not weakly correlated systems and cannot be compared to overdoped cuprates with reduced electronic correlations but rather need to be placed in the family of multiorbital correlated systems similarly to heavy fermion materials \cite{yang_evidence_2009,monney_resonant_2016}.

\subsection*{X-Ray Emission Spectroscopy}
To complement our measurements of the spin excitations and assess the local magnetism of \BFAP, we employed X-ray Emission Spectroscopy (XES) - a classical technique that has been established as a sensitive probe of the local magnetic moment ($\mu_{bare}$) \cite{gretarsson_spin-state_2013,gretarsson_revealing_2011,glatzel_high_2005,vanko_probing_2006,pelliciari_local_2017,yamamoto_origin_2016,lafuerza_evidence_2017,pelliciari_magnetic_2017}. XES is sensitive to the local fluctuating magnetic moment and does not require a net magnetization or ordering, such as X-ray Magnetic Circular Dichroism (XMCD) or neutron diffraction, but detects directly the paramagnetic moment \cite{glatzel_high_2005,vanko_probing_2006,yamamoto_origin_2016}. In this technique, a photon (h$\nu=$7140~eV) excites an Fe-\textit{1s} core-electron into the continuum, creating a core-hole which is filled by a Fe-\textit{3p} electron with the consequent emission of a photon (h$\nu=$7040-7065~eV). The final Fe-\textit{3p}$^5$ state has a wave function partly overlapped with the Fe-\textit{3d} orbitals, and is, thereby, affected by the spin of the valence band through the exchange interaction \cite{vilmercati_itinerant_2012,gretarsson_spin-state_2013,gretarsson_revealing_2011,glatzel_high_2005,vanko_probing_2006,pelliciari_local_2017,lafuerza_evidence_2017,pelliciari_magnetic_2017}. The femtosecond timescale of this technique allows for the measurement of fast fluctuations of $\mu_{bare}$ \cite{mannella_magnetic_2014,gretarsson_spin-state_2013,gretarsson_revealing_2011,vilmercati_itinerant_2012}, and by means of normalization and calibration [carried out with FeCrAs (0~$\mu_B$) and \BFA (set arbitrarily to one to define a relative scale)] $\mu_{bare}$ can be determined \cite{gretarsson_spin-state_2013,gretarsson_revealing_2011,vilmercati_itinerant_2012}. 

\begin{figure}
\includegraphics[scale=0.55,trim={0 260 220 0},clip]{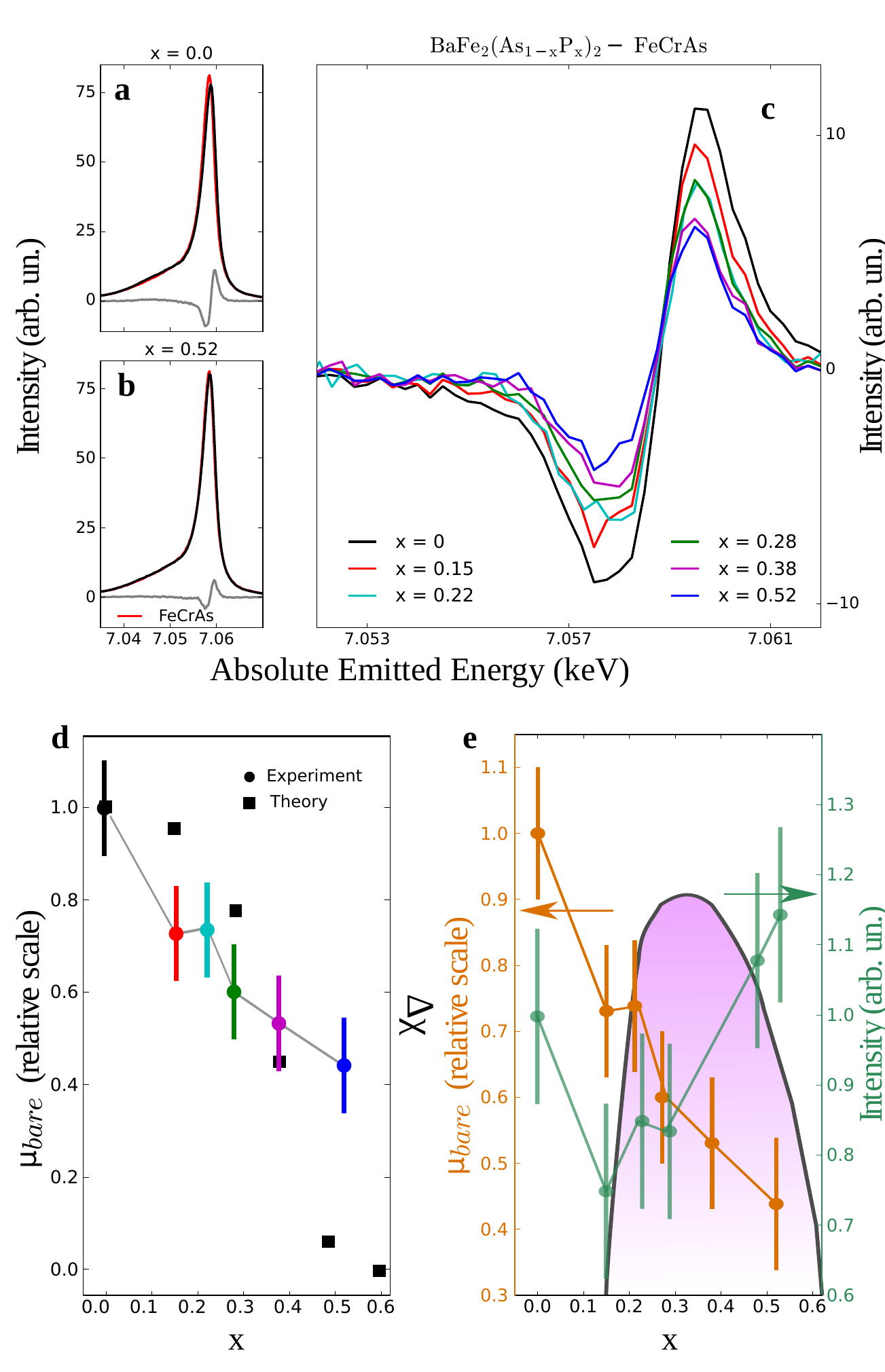}
\caption{\label{fig:fig4} \textbf{Fe-K$_\beta$ X-ray Emission Spectroscopy (XES) and difference spectra for \BFAP}. K$_\beta$ XES for $x=$0.00, and 0.52 (\textbf{a,b} as black lines) and reference spectrum of FeCrAs (\textbf{a,b} as red lines). The Integrated Area Difference (IAD) is displayed as grey line. \textbf{c} Summary of IADs for all the samples. \textbf{d} Values of local magnetic moment extracted from XES compared with theoretical estimates of the Fermi surface (FS) nesting strength. The nesting strength is calculated integrating the static susceptibility  at the nesting wavevector. The error bars on the values of~$\mu_{bare}$  are systematic due to the normalization and subtraction steps, thus all the values have an error of $\pm$0.1 with respect to the values assumed by the parent compound.  \textbf{e} Summary of local magnetic moment from XES (orange) and intensity of spin excitations (green) integrating the fitting from the Resonant Inelastic X-ray Scattering (RIXS) spectra at high momenta (three highest momenta) along (H,0) and (H,H) as a function of doping across the phase diagram of \BFAP. In order to visualize the reciprocal redistribution of local magnetic moments and spin excitations we represent the superconducting dome as a purple shaded area in the background. The error bars on the intensity of the spin excitations have been quantified in $\pm$0.1 with respect to the value of the parent compound. The fitting error are actually smaller but the sum procedure at different q points produces a propagation of the errors, so we set a slightly higher upper boundary.}
\end{figure}

In Fig.~\ref{fig:fig4}\textbf{a,b}, we show XES spectra for \BFA, and \BFAP (x=0.52), and FeCrAs and the respective difference spectra. Clearly, a gradual decrease of $\mu_{bare}$ is inferred from the difference spectra depicted in Fig.~\ref{fig:fig4}\textbf{c}. The values of $\mu_{bare}$, presented in Fig.~\ref{fig:fig4}\textbf{d,e} are continuously reduced, (for example $\mu_{bare}=$ 1.0$\pm 0.1$ for $x=$0.00, $\mu_{bare}=0.6\pm0.1$ for $x=$0.28, and $\mu_{bare}=0.4\pm 0.1$ for $x=$0.52), but not fully quenched by doping, despite the complete disappearance of the ordered magnetic moment observed by neutron scattering \cite{hu_structural_2015,allred_coincident_2014}. This evidence is remarkable in light of the constancy of the Fe oxidation state observed in XAS (see Supplementary Fig. 2) with isovalent doping driving the antiferromagnetically long range ordered \BFA~into a paramagnetic phase.

We compare the doping dependence of the local moment with the strength of the Fermi surface (FS) nesting at the antiferromagnetic wavevector. In Fig.\ref{fig:fig4}\textbf{d}, the theoretical data is presented as $\Delta\chi$(x) = $\chi$(x) - $\chi$(0.6), where  $\chi$(x) is the computed static susceptibility at the antiferromagnetic wavevector at doping x [we normalize $\Delta\chi$ (x=0) =1]. Note that this comparison is only qualitative, and a self-consistent estimation of the static magnetic moment is computationally expensive. Similar doping dependence of the magnetic instability arising from the FS nesting and the observed local moment indicates an electronic mechanism of the magnetic ground state in this system.

\section*{Discussion}
Our XES results are puzzling and assuming spectral weight sum rules, there has to be spectral weight transfer from localized moments into spin excitations as probed by RIXS. In Fig.~\ref{fig:fig4}\textbf{e}, we report the integrated spectral weight of the spin excitations determined from our RIXS spectra. Within error bars, the spin excitations intensities detected in the portion of BZ pertinent to RIXS possibly show gradual transfer of spectral weight upon doping. However, additional enhancement of spectral weight in different regions of the BZ could be observed. On this aspect, further combined studies of RIXS and high energy INS as a function of doping are required to shed light on the redistribution of spectral weight in the complete BZ.
In Fig.~\ref{fig:fig4}\textbf{e}, we observe the gradual transfer of spectral weight from local magnetic moments into collective spin excitations across the superconducting dome. This spectral redistribution signals that SC is optimized by a balance of these types of interactions.

In summary, we unveiled the persistence and hardening of magnetic excitations along the superconducting dome of \BFAP with concomitant decrease of local magnetic moments.  The spectral weight of the spin excitations observed with RIXS slightly increases corroborating that SC emerges from a balance of localized and itinerant electronic interactions. The experimental results agree well with intermediate coupling calculations placing the FePns in the family of multi-orbital correlated systems and uncovering clearly the mixed electronic and magnetic interactions present with optimal superconductivity.

\section*{Methods} 
\subsection*{Sample preparations}
Single crystals of \BFAP were grown either by stoichiometric melt \cite{kasahara_evolution_2010} or Ba$_2$As$_3$ / Ba$_2$P$_3$ self-flux method \cite{nakajima_growth_2012}. The samples were characterized either with resistivity or magnetization measurements. For $x=0.15$, 0.22, 0.27, and 0.48, crystals grown by the stoichiometric melt technique were used for the Resonant Inelastic X-ray Scattering (RIXS) experiments, while for $x=0.38$ and 0.52 we use the crystals grown by the self-flux technique. Supplementary Figure 1 shows the temperature dependence of the resistivity on crystals selected from the same batches. For $x>0.3$ the Spin Density Wave (SDW) long range order is completely suppressed \cite{kasahara_evolution_2010,nakajima_growth_2012}. 

\subsection*{X-ray Absorption Spectroscopy and Resonant Inelastic X-ray Scattering}
X-ray Absorption Spectra (XAS) and RIXS experiments were performed at the ADRESS beamline of the Swiss Light Source, Paul Scherrer Institut, Villigen PSI, Switzerland \cite{strocov_high-resolution_2010,ghiringhelli_saxes_2006}. In agreement with previous works \cite{zhou_persistent_2013, pelliciari_presence_2016, pelliciari_intralayer_2016, hancock_evidence_2010,pelliciari_local_2017}, the samples were mounted with the \textit{ab} plane perpendicular to the scattering plane and the \textit{c} axis lying in it (sketch in Supplementary Figure 2) and post-cleaved \textit{in-situ} at a pressure better than 2.0$\times$10$^{-10}$ mbar. The reciprocal space directions studied are (0, 0)$\rightarrow$(1, 0) and (0, 0)$\rightarrow$(1, 1) according to the orthorhombic unfolded crystallographic notation. The values of in-plane momentum transferred are expressed as relative lattice units (RLU) (q$_{//}$a / 2$\pi$). We use the convention of 1 Fe per unit cell. All the measurements were carried out at 13~K by cooling the manipulator with liquid helium. XAS spectra were measured in both Total Fluorescence Yield (TFY) and Total Electron Yield (TEY) modes. No difference was observed between TEY and TFY which indicates the sample integrity across the thickness probed in our experiment. We measured Fe L$_{2,3}$ XAS spectra for all samples at 15 degrees of incidence angle relative to the sample surface. All the XAS spectra are reported in Supplementary Figure 3 and display the constancy of the iron oxidation state. There are small spectral differences at around 710 eV that are possibly due to a different covalency between the FeAs and FeP.

The RIXS spectrometer was set to a scattering angle of 130 degrees and the incidence angle on the samples surface was varied to change the in-plane momentum transferred (q$_{//}$) from (0, 0) to (0.44, 0) and from (0, 0) to (0.31, 0.31). All RIXS measurements in the present paper were recorded in grazing incidence configuration as depicted in Supplementary Figure 2. The zero energy loss of our RIXS spectra was determined by measuring spectra in $\sigma$ polarization. The total energy resolution was measured employing the elastic scattering of carbon-filled acrylic tape and is around 110~meV. 

RIXS spectra were normalized to unity and the main emission line was fitted according to Ref.~\citep{zhou_persistent_2013,pelliciari_presence_2016,pelliciari_intralayer_2016,hancock_evidence_2010,pelliciari_local_2017} employing the following formulas:

\begin{equation}
I_{fit} = (\beta x^{2} + \alpha x + c )  \cdot(1 - g_{\gamma}) + I_{0}\exp(ax) \cdot g_{\gamma} + G \nonumber 
\end{equation} 
\textrm{with}\\

\begin{equation}
g_{\gamma} =  \left(\exp \left(\frac{x + \omega^{\ast}} {\Gamma}\right) + 1\right)^{-1} \nonumber 
\end{equation} 
\textrm{and}\\
\begin{equation}
G = \frac{A}{\sigma \sqrt{2\pi}} \exp\left(\frac{(x + x_{0})^2} {2\sigma^2}\right) \nonumber 
\end{equation} 
\\
In the first formula, the first part is a 2$^{nd}$ order polynomial function describing the emission line at low energy loss, the second part is an exponential decay describing the emission line at high energy loss. The two behaviours are swapped into each other by the g$_\gamma$ term. The third term is a gaussian curve (G) observed at around -4.2~eV energy loss.
An exemplary fitting of the full RIXS spectra is drawn in Supplementary Figure 4 for \BFA at (0.44, 0.0).

In Supplementary Figure 5, we show RIXS spectra over an extended range of energy loss. Overall, the three spectra resemble a metallic system with the lineshape not being affected by doping. Due to changes in the band structure upon doping, small energy shifts and changes in the RIXS intensity are detected. 

We complemented the momentum dependence along (0, 0)$\rightarrow$(1, 0) with (0, 0)$\rightarrow$(1, 1) and display the raw data for all the dopings and momenta with the fitting in Supplementary Figure 6.

\subsection*{K$_\beta$ Emission Spectroscopy}
We performed the Fe-K$_\beta$ x-ray emission (XES) experiments at BL11XU of SPring-8, Japan. The incoming beam was monochromatized by a Si~(111) double-crystal and a Si~(400) secondary channel-cut crystal. The energy was calibrated by measuring x-ray absorption of a Fe foil and set to 7.140~keV with $\pi$ polarization. We employed three spherical diced Ge~(620) analyzers and a detector in Rowland geometry at 2~m distance. The total combined resolution was about 400~meV estimated from Full Width Half Maximum (FWHM) of the elastic line. We scanned the absolute emission energy between 7.02~keV and 7.08~keV and normalized the intensity by the incident flux monitored by an ionization chamber. We carried out the measurements at 15~K. The experimental geometry of the experiment is depicted in Supplementary Figure 7.
The determination of $\mu_{bare}$ were performed employing the integrated absolute difference (IAD) method described by Vanko' et al.~\cite{vanko_probing_2006}. The areas of the XES spectra were normalized to unity as shown in Supplementary Figure 8\textbf{a-f}. To overcome problems of alignment of the absolute energy between different samples, we estimated the center of mass of the spectra and aligned the energy to have the same center of mass as described by Glatzel et al.~\cite{glatzel_high_2005}. Then the difference with the spectrum of the FeCrAs reference was taken as displayed in Supplementary Figure 8\textbf{a-f} and summarized in Supplementary Figure 8\textbf{g}. The integration of this difference gives the IAD, which is directly proportional to $\mu_{bare}$ \cite{vanko_probing_2006}. 

\subsection*{Doping dependent band structure}
Two extreme doping regimes of the BaFe$_2$(As$_{1-x}$P$_x$)$_2$ samples, namely BaFe$_2$As$_2$ ($x=0.00$) and BaFe$_2$P$_2$ ($x=1.00$) have been studied within density functional theory (DFT) \cite{graser_near-degeneracy_2009}. Also accurate tight-binding models are available to reproduce the DFT band structure, consistent with Angle Resolved Photoemission Spectroscopy (ARPES) data (after including renormalization effects). However, P-doping effects on the As site, which does not change the carrier concentration, is neither trivial to accurately calculate within the DFT framework, nor within a simple rigid band shift technique which works reasonably well for electron and hole doping cases \cite{richard_fe-based_2011}. Since P atom is smaller in size compared to As atoms, P doping is expected to decrease the lattice volume. Indeed, both x-ray \cite{kasahara_evolution_2010}, and neutron \cite{allred_coincident_2014} diffraction analysis revealed that all three lattice constants as well as the pnictogen atomic coordinates in the unit cell (z$_{Pn}$) and the pnictogen height from the Fe plane (h$_{Pn}$) decrease monotonically with P doping. Among them, the \textit{c}-axis lattice constant decreases drastically from $\approx$13 \AA ~at $x=0.00$ to $\approx$12.4 \AA ~at $x=1.00$. This explains why the three-dimensionality of the sample (i.e. k$_z$ dispersion) increases considerably with doping, as consistently demonstrated in both ARPES \cite{yoshida_two-dimensional_2011} as well as in de Haas-van Alphen measurements \cite{arnold_nesting_2011,analytis_enhanced_2010}. It is shown that throughout the entire doping range, the Fermi surface (FS) topology remains very similar and consists of hole pockets at the $\Gamma$-point and electron pockets at the M-point. At $x=0.00$, there are three concentric hole pockets, which reduce to only two remaining hole pockets at $x=1.00$, with the smaller pocket disappearing at some intermediate doping. The electron-like FS consists of two elliptic electron pockets interpenetrating to each other \cite{shibauchi_quantum_2014}. Among all the FS pockets, the outermost hole-pocket shows most dominant and drastic changes with doping: It is very close to cylindrical as a function of k$_z$ at $x=0.00$, but gradually becomes corrugated with increasing doping. The other pockets acquire comparatively less k$_z$ dispersion \cite{yoshida_two-dimensional_2011,analytis_enhanced_2010,arnold_nesting_2011}. This could be quantified by calculating the nesting strength via the integration of the static susceptibility at the nesting wave-vector. The nesting strength has been shown to decrease upon doping \cite{arnold_nesting_2011,richard_fe-based_2011,shibauchi_quantum_2014}. Projected orbital symmetries on each FS reveal that the hole pockets are dominated by the \textit{d}$_{xz}$ / \textit{d}$_{yz}$ orbitals, while the electron pockets are made of hybridization between the \textit{d}$_{xz}$ / \textit{d}$_{yz}$ orbitals with the \textit{d}$_{xy}$ orbitals. The \textit{d}$_{z^2}$ orbital has little contribution at $x=0.00$, while it grows with doping. \textit{d}$_{z^2}$ basically contributes to the strongly k$_z$ dispersive parts of the outermost hole-pocket \cite{arnold_nesting_2011}.

Based on the aforementioned experimental observations on the band structure evolution with doping, we construct an effective five-orbital tight-binding (TB) model to reproduce the low-energy dispersion and FS topology across the entire phase diagram. The 2 Fe unit cell incorporates two Fe sublattices producing a 10 band model. We start with a five orbital TB model as derived in Ref.~\cite{graser_near-degeneracy_2009} for BaFe$_2$As$_2$. The intra-orbital dispersions are defined by $\xi^i_k= \epsilon^i_k+\Delta^i-\epsilon_F $, where \textit{k} is the crystal momentum and $i=1-5$ is the orbital index. $\epsilon^i_k$ is the momentum dependent part of the dispersion which arises from the nearest, next, and higher neighbour hoppings between the same orbitals, $\Delta^i$ is the corresponding on-site potentials, and $\epsilon_F$ is the chemical potential. As the unit cell volume decreases monotonically with P doping, it is expected that the electron hopping amplitude increases monotonically with doping. We model this effect by a simple renormalization factor $\lambda$ as $\epsilon^i_k\rightarrow\lambda\epsilon^i_k$, where $\lambda$ increases with doping. Similarly, due to the monotonic increase of the pnictogen coordinate and the height (z$_{Pn}$, h$_{Pn}$), the on-site potential also changes. Interestingly, we find that an orbital dependent modification of the on-site potential is required to properly reproduce the experimental behaviour of the k$_z$ dispersion and FS changes. We set $\Delta^i\rightarrow\Delta^i+\delta$ for \textit{d}$_{xz}$ / \textit{d}$_{yz}$ orbitals, $\Delta^i\rightarrow\Delta^i-\delta$ for \textit{d}$_{x^2-y^2}$ and \textit{d}$_z^2$ orbitals, $\Delta^i\rightarrow\Delta^i-2\delta$ for \textit{d}$_{xy}$ orbital, where $\delta(x)=0.08x$, with $x$ being the doping value. The chemical potential is adjusted at each doping to keep the same number of electrons. Such a model was also invoked earlier to model the FS topological transition in KFe$_2$Se$_2$ under pressure \cite{das_origin_2013}. The corresponding FS topologies at three representative dopings are shown in Supplementary Figure 9. 

\section*{Authors contribution}
Th. Sc. conceived the project. J. P., K.I., Y.H., and Th. Sc. conducted XES experiments. J. P., Y. H., M. D., P. O. V., X. L., and Th. Sc. performed RIXS experiments with the assistance of V. N. S.. S. K., Y. M., and Ta. Sh. prepared and characterized P-doped \BFA. L. X., X. W., and C. J. prepared the reference sample for XES. T. D. performed the calculations. J. P., K. I., T.D., and Th. Sc. together planned the project phases. J. P., T.D., and Th. Sc. wrote the manuscript with input from all the authors.

\section*{Acknowledgements}
J.P. and T.S. acknowledge financial support through the Dysenos AG by Kabelwerke Brugg AG Holding, Fachhochschule Nordwestschweiz, and the Paul Scherrer Institut. J. P. also acknowledges financial support by the Swiss National Science Foundation Early Postdoc Mobility fellowship project number P2FRP2$\_$171824 and PostDoc Mobility project number P400P2$\_$180744. We acknowledge Y. Shimizu for the support during the experiments at SPring-8 and D. Casa for fabrication of the Ge (620) analyzers. The Fe-K$_\beta$ emission experiments were performed at BL11XU of SPring-8 with the approval of the Japan Synchrotron Radiation Research Institute (JASRI) (Proposals No. 2014A3502 and 2014B3502). RIXS experiments have been performed at the ADRESS beamline of the Swiss Light Source at Paul Scherrer Institut. Part of this research has been funded by the Swiss National Science Foundation through the Sinergia network Mott Physics Beyond the Heisenberg (MPBH) model (SNSF Research grants numbers CRSII2$\_$141962 and CRSII2$\_$160765$\/$1) and the D-A-CH program (SNSF Research Grant No. 200021L$\_$141325). The research leading to these results has received funding from the European Community's Seventh Framework Programme (FP7/2007-2013) under Grant Agreement No. 290605 (COFUND: PSIFELLOW). Work in Japan was supported by Grant-in-Aids for Scientific Research (KAKENHI) from Japan Society for the Promotion of Science (JSPS), and by the `Topological Material Science' Grant-in-Aid for Scientific Research on Innovative Areas from the Ministry of Education, Culture, Sports, Science and Technology (MEXT) of Japan. The work at IOP-CAS is supported by NSF and MOST through research projects.

\section*{Competing Interests} 
The authors declare no competing interests.

\section*{Data Availability}
Data that support the findings of this study are available upon reasonable request from the corresponding authors.

\bibliography{MyLibrary_v8}{}

\begin{thebibliography}{10}
\expandafter\ifx\csname url\endcsname\relax
  \def\url#1{\texttt{#1}}\fi
\expandafter\ifx\csname urlprefix\endcsname\relax\def\urlprefix{URL }\fi
\providecommand{\bibinfo}[2]{#2}
\providecommand{\eprint}[2][]{\url{#2}}

\bibitem{johnston_puzzle_2010}
\bibinfo{author}{Johnston, D.~C.}
\newblock \bibinfo{title}{The puzzle of high temperature superconductivity in
  layered iron pnictides and chalcogenides}.
\newblock \emph{\bibinfo{journal}{Advances in Physics}}
  \textbf{\bibinfo{volume}{59}}, \bibinfo{pages}{803--1061}
  (\bibinfo{year}{2010}).
\newblock \urlprefix\url{http://dx.doi.org/10.1080/00018732.2010.513480}.

\bibitem{stewart_superconductivity_2011}
\bibinfo{author}{Stewart, G.~R.}
\newblock \bibinfo{title}{Superconductivity in iron compounds}.
\newblock \emph{\bibinfo{journal}{Reviews of Modern Physics}}
  \textbf{\bibinfo{volume}{83}}, \bibinfo{pages}{1589--1652}
  (\bibinfo{year}{2011}).
\newblock \urlprefix\url{http://link.aps.org/doi/10.1103/RevModPhys.83.1589}.

\bibitem{chubukov_pairing_2012}
\bibinfo{author}{Chubukov, A.}
\newblock \bibinfo{title}{Pairing {Mechanism} in {Fe}-{Based}
  {Superconductors}}.
\newblock \emph{\bibinfo{journal}{Annual Review of Condensed Matter Physics}}
  \textbf{\bibinfo{volume}{3}}, \bibinfo{pages}{57--92} (\bibinfo{year}{2012}).
\newblock
  \urlprefix\url{http://dx.doi.org/10.1146/annurev-conmatphys-020911-125055}.

\bibitem{scalapino_common_2012}
\bibinfo{author}{Scalapino, D.~J.}
\newblock \bibinfo{title}{A common thread: {The} pairing interaction for
  unconventional superconductors}.
\newblock \emph{\bibinfo{journal}{Reviews of Modern Physics}}
  \textbf{\bibinfo{volume}{84}}, \bibinfo{pages}{1383--1417}
  (\bibinfo{year}{2012}).
\newblock \urlprefix\url{http://link.aps.org/doi/10.1103/RevModPhys.84.1383}.

\bibitem{mannella_magnetic_2014}
\bibinfo{author}{Mannella, N.}
\newblock \bibinfo{title}{The magnetic moment enigma in {Fe}-based high
  temperature superconductors}.
\newblock \emph{\bibinfo{journal}{Journal of Physics: Condensed Matter}}
  \textbf{\bibinfo{volume}{26}}, \bibinfo{pages}{473202}
  (\bibinfo{year}{2014}).
\newblock \urlprefix\url{http://stacks.iop.org/0953-8984/26/i=47/a=473202}.

\bibitem{inosov_spin_nodate}
\bibinfo{author}{Inosov, D.~S.}
\newblock \bibinfo{title}{Spin fluctuations in iron pnictides and
  chalcogenides: {From} antiferromagnetism to superconductivity}.
\newblock \emph{\bibinfo{journal}{Comptes Rendus Physique}}
  \textbf{\bibinfo{volume}{17}}, \bibinfo{pages}{60} (\bibinfo{year}{2016}).
\newblock
  \urlprefix\url{http://www.sciencedirect.com/science/article/pii/S1631070515000523}.

\bibitem{tranquada_superconductivity_2014}
\bibinfo{author}{Tranquada, J.~M.}, \bibinfo{author}{Xu, G.} \&
  \bibinfo{author}{Zaliznyak, I.~A.}
\newblock \bibinfo{title}{Superconductivity, antiferromagnetism, and neutron
  scattering}.
\newblock \emph{\bibinfo{journal}{Journal of Magnetism and Magnetic Materials}}
  \textbf{\bibinfo{volume}{350}}, \bibinfo{pages}{148--160}
  (\bibinfo{year}{2014}).
\newblock
  \urlprefix\url{http://www.sciencedirect.com/science/article/pii/S0304885313006884}.

\bibitem{dai_antiferromagnetic_2015}
\bibinfo{author}{Dai, P.}
\newblock \bibinfo{title}{Antiferromagnetic order and spin dynamics in
  iron-based superconductors}.
\newblock \emph{\bibinfo{journal}{Reviews of Modern Physics}}
  \textbf{\bibinfo{volume}{87}}, \bibinfo{pages}{855--896}
  (\bibinfo{year}{2015}).
\newblock \urlprefix\url{http://link.aps.org/doi/10.1103/RevModPhys.87.855}.

\bibitem{dai_magnetism_2012}
\bibinfo{author}{Dai, P.}, \bibinfo{author}{Hu, J.} \&
  \bibinfo{author}{Dagotto, E.}
\newblock \bibinfo{title}{Magnetism and its microscopic origin in iron-based
  high-temperature superconductors}.
\newblock \emph{\bibinfo{journal}{Nature Physics}}
  \textbf{\bibinfo{volume}{8}}, \bibinfo{pages}{709--718}
  (\bibinfo{year}{2012}).
\newblock
  \urlprefix\url{http://www.nature.com/nphys/journal/v8/n10/abs/nphys2438.html}.

\bibitem{lumsden_magnetism_2010}
\bibinfo{author}{Lumsden, M.~D.} \& \bibinfo{author}{Christianson, A.~D.}
\newblock \bibinfo{title}{Magnetism in {Fe}-based superconductors}.
\newblock \emph{\bibinfo{journal}{Journal of Physics: Condensed Matter}}
  \textbf{\bibinfo{volume}{22}}, \bibinfo{pages}{203203}
  (\bibinfo{year}{2010}).
\newblock \urlprefix\url{http://stacks.iop.org/0953-8984/22/i=20/a=203203}.

\bibitem{shibauchi_quantum_2014}
\bibinfo{author}{Shibauchi, T.}, \bibinfo{author}{Carrington, A.} \&
  \bibinfo{author}{Matsuda, Y.}
\newblock \bibinfo{title}{A {Quantum} {Critical} {Point} {Lying} {Beneath} the
  {Superconducting} {Dome} in {Iron} {Pnictides}}.
\newblock \emph{\bibinfo{journal}{Annual Review of Condensed Matter Physics}}
  \textbf{\bibinfo{volume}{5}}, \bibinfo{pages}{113--135}
  (\bibinfo{year}{2014}).
\newblock
  \urlprefix\url{http://dx.doi.org/10.1146/annurev-conmatphys-031113-133921}.

\bibitem{lee_doping_2006}
\bibinfo{author}{Lee, P.~A.}, \bibinfo{author}{Nagaosa, N.} \&
  \bibinfo{author}{Wen, X.-G.}
\newblock \bibinfo{title}{Doping a {Mott} insulator: {Physics} of
  high-temperature superconductivity}.
\newblock \emph{\bibinfo{journal}{Reviews of Modern Physics}}
  \textbf{\bibinfo{volume}{78}}, \bibinfo{pages}{17--85}
  (\bibinfo{year}{2006}).
\newblock \urlprefix\url{http://link.aps.org/doi/10.1103/RevModPhys.78.17}.

\bibitem{yin_kinetic_2011}
\bibinfo{author}{Yin, Z.~P.}, \bibinfo{author}{Haule, K.} \&
  \bibinfo{author}{Kotliar, G.}
\newblock \bibinfo{title}{Kinetic frustration and the nature of the magnetic
  and paramagnetic states in iron pnictides and iron chalcogenides}.
\newblock \emph{\bibinfo{journal}{Nature Materials}}
  \textbf{\bibinfo{volume}{10}}, \bibinfo{pages}{932--935}
  (\bibinfo{year}{2011}).
\newblock
  \urlprefix\url{http://www.nature.com/nmat/journal/v10/n12/full/nmat3120.html}.

\bibitem{de_medici_orbital-selective_2009}
\bibinfo{author}{de’ Medici, L.}, \bibinfo{author}{Hassan, S.~R.},
  \bibinfo{author}{Capone, M.} \& \bibinfo{author}{Dai, X.}
\newblock \bibinfo{title}{Orbital-{Selective} {Mott} {Transition} out of {Band}
  {Degeneracy} {Lifting}}.
\newblock \emph{\bibinfo{journal}{Physical Review Letters}}
  \textbf{\bibinfo{volume}{102}}, \bibinfo{pages}{126401}
  (\bibinfo{year}{2009}).
\newblock
  \urlprefix\url{https://link.aps.org/doi/10.1103/PhysRevLett.102.126401}.

\bibitem{chubukov_itinerant_2015}
\bibinfo{author}{Chubukov, A.~V.}
\newblock \bibinfo{title}{Itinerant electron scenario for {Fe}-based
  superconductors}.
\newblock \emph{\bibinfo{journal}{Springer Series in Materials Science}}
  \textbf{\bibinfo{volume}{211}}, \bibinfo{pages}{255--329}
  (\bibinfo{year}{2015}).
\newblock \urlprefix\url{http://xxx.tau.ac.il/abs/1507.03856}.

\bibitem{yang_evidence_2009}
\bibinfo{author}{Yang, W.~L.} \emph{et~al.}
\newblock \bibinfo{title}{Evidence for weak electronic correlations in iron
  pnictides}.
\newblock \emph{\bibinfo{journal}{Physical Review B}}
  \textbf{\bibinfo{volume}{80}}, \bibinfo{pages}{014508}
  (\bibinfo{year}{2009}).
\newblock \urlprefix\url{http://link.aps.org/doi/10.1103/PhysRevB.80.014508}.

\bibitem{zhou_persistent_2013}
\bibinfo{author}{Zhou, K.-J.} \emph{et~al.}
\newblock \bibinfo{title}{Persistent high-energy spin excitations in
  iron-pnictide superconductors}.
\newblock \emph{\bibinfo{journal}{Nature Communications}}
  \textbf{\bibinfo{volume}{4}}, \bibinfo{pages}{1470} (\bibinfo{year}{2013}).
\newblock
  \urlprefix\url{http://www.nature.com/ncomms/journal/v4/n2/full/ncomms2428.html}.

\bibitem{pelliciari_presence_2016}
\bibinfo{author}{Pelliciari, J.} \emph{et~al.}
\newblock \bibinfo{title}{Presence of magnetic excitations in {SmFeAsO}}.
\newblock \emph{\bibinfo{journal}{Applied Physics Letters}}
  \textbf{\bibinfo{volume}{109}}, \bibinfo{pages}{122601}
  (\bibinfo{year}{2016}).
\newblock
  \urlprefix\url{http://scitation.aip.org/content/aip/journal/apl/109/12/10.1063/1.4962966}.

\bibitem{pelliciari_local_2017}
\bibinfo{author}{Pelliciari, J.} \emph{et~al.}
\newblock \bibinfo{title}{Local and collective magnetism of {EuFe$_2$As$_2$}}.
\newblock \emph{\bibinfo{journal}{Physical Review B}}
  \textbf{\bibinfo{volume}{95}}, \bibinfo{pages}{115152}
  (\bibinfo{year}{2017}).
\newblock \urlprefix\url{https://link.aps.org/doi/10.1103/PhysRevB.95.115152}.

\bibitem{pelliciari_intralayer_2016}
\bibinfo{author}{Pelliciari, J.} \emph{et~al.}
\newblock \bibinfo{title}{Intralayer doping effects on the high-energy magnetic
  correlations in {NaFeAs}}.
\newblock \emph{\bibinfo{journal}{Physical Review B}}
  \textbf{\bibinfo{volume}{93}}, \bibinfo{pages}{134515}
  (\bibinfo{year}{2016}).
\newblock \urlprefix\url{http://link.aps.org/doi/10.1103/PhysRevB.93.134515}.

\bibitem{rahn_paramagnon_2019}
\bibinfo{author}{Rahn, M.~C.} \emph{et~al.}
\newblock \bibinfo{title}{Paramagnon dispersion in $\beta$-{FeSe} observed by
  {Fe} {L}-edge resonant inelastic x-ray scattering}.
\newblock \emph{\bibinfo{journal}{Physical Review B}}
  \textbf{\bibinfo{volume}{99}}, \bibinfo{pages}{014505}
  (\bibinfo{year}{2019}).
\newblock \urlprefix\url{https://link.aps.org/doi/10.1103/PhysRevB.99.014505}.

\bibitem{garcia_anisotropic_2019}
\bibinfo{author}{Garcia, F.~A.} \emph{et~al.}
\newblock \bibinfo{title}{Anisotropic magnetic excitations and incipient {Neel}
  order in {Ba(Fe$_{1-x}$Mn$_x$)$_2$As$_2$}}.
\newblock \emph{\bibinfo{journal}{Physical Review B}}
  \textbf{\bibinfo{volume}{99}}, \bibinfo{pages}{115118}
  (\bibinfo{year}{2019}).
\newblock \urlprefix\url{https://link.aps.org/doi/10.1103/PhysRevB.99.115118}.

\bibitem{ament_resonant_2011}
\bibinfo{author}{Ament, L. J.~P.}, \bibinfo{author}{van Veenendaal, M.},
  \bibinfo{author}{Devereaux, T.~P.}, \bibinfo{author}{Hill, J.~P.} \&
  \bibinfo{author}{van~den Brink, J.}
\newblock \bibinfo{title}{Resonant inelastic x-ray scattering studies of
  elementary excitations}.
\newblock \emph{\bibinfo{journal}{Reviews of Modern Physics}}
  \textbf{\bibinfo{volume}{83}}, \bibinfo{pages}{705--767}
  (\bibinfo{year}{2011}).
\newblock \urlprefix\url{http://link.aps.org/doi/10.1103/RevModPhys.83.705}.

\bibitem{jia_using_2016}
\bibinfo{author}{Jia, C.}, \bibinfo{author}{Wohlfeld, K.},
  \bibinfo{author}{Wang, Y.}, \bibinfo{author}{Moritz, B.} \&
  \bibinfo{author}{Devereaux, T.~P.}
\newblock \bibinfo{title}{Using {RIXS} to {Uncover} {Elementary} {Charge} and
  {Spin} {Excitations}}.
\newblock \emph{\bibinfo{journal}{Physical Review X}}
  \textbf{\bibinfo{volume}{6}}, \bibinfo{pages}{021020} (\bibinfo{year}{2016}).
\newblock \urlprefix\url{http://link.aps.org/doi/10.1103/PhysRevX.6.021020}.

\bibitem{tohyama_enhanced_2015}
\bibinfo{author}{Tohyama, T.}, \bibinfo{author}{Tsutsui, K.},
  \bibinfo{author}{Mori, M.}, \bibinfo{author}{Sota, S.} \&
  \bibinfo{author}{Yunoki, S.}
\newblock \bibinfo{title}{Enhanced charge excitations in electron-doped
  cuprates by resonant inelastic x-ray scattering}.
\newblock \emph{\bibinfo{journal}{Physical Review B}}
  \textbf{\bibinfo{volume}{92}}, \bibinfo{pages}{014515}
  (\bibinfo{year}{2015}).
\newblock \urlprefix\url{https://link.aps.org/doi/10.1103/PhysRevB.92.014515}.

\bibitem{hu_spin_2016}
\bibinfo{author}{Hu, D.} \emph{et~al.}
\newblock \bibinfo{title}{Spin excitations in optimally {P}-doped
  {BaFe$_2$(As$_{1-x}$P$_x$)$_2$} superconductor}.
\newblock \emph{\bibinfo{journal}{Physical Review B}}
  \textbf{\bibinfo{volume}{94}}, \bibinfo{pages}{094504}
  (\bibinfo{year}{2016}).
\newblock \urlprefix\url{http://link.aps.org/doi/10.1103/PhysRevB.94.094504}.

\bibitem{nomura_resonant_2016}
\bibinfo{author}{Nomura, T.} \emph{et~al.}
\newblock \bibinfo{title}{Resonant inelastic x-ray scattering study of
  entangled spin-orbital excitations in superconducting {PrFeAsO$_{0.7}$}}.
\newblock \emph{\bibinfo{journal}{Physical Review B}}
  \textbf{\bibinfo{volume}{94}}, \bibinfo{pages}{035134}
  (\bibinfo{year}{2016}).
\newblock \urlprefix\url{http://link.aps.org/doi/10.1103/PhysRevB.94.035134}.

\bibitem{kaneshita_spin_2011}
\bibinfo{author}{Kaneshita, E.}, \bibinfo{author}{Tsutsui, K.} \&
  \bibinfo{author}{Tohyama, T.}
\newblock \bibinfo{title}{Spin and orbital characters of excitations in iron
  arsenide superconductors revealed by simulated resonant inelastic x-ray
  scattering}.
\newblock \emph{\bibinfo{journal}{Physical Review B}}
  \textbf{\bibinfo{volume}{84}}, \bibinfo{pages}{020511}
  (\bibinfo{year}{2011}).
\newblock \urlprefix\url{http://link.aps.org/doi/10.1103/PhysRevB.84.020511}.

\bibitem{wang_doping_2013}
\bibinfo{author}{Wang, M.} \emph{et~al.}
\newblock \bibinfo{title}{Doping dependence of spin excitations and its
  correlations with high-temperature superconductivity in iron pnictides}.
\newblock \emph{\bibinfo{journal}{Nature Communications}}
  \textbf{\bibinfo{volume}{4}} (\bibinfo{year}{2013}).
\newblock
  \urlprefix\url{http://www.nature.com/ncomms/2013/131204/ncomms3874/full/ncomms3874.html}.

\bibitem{eilers_strain-driven_2016}
\bibinfo{author}{Eilers, F.} \emph{et~al.}
\newblock \bibinfo{title}{Strain-{Driven} {Approach} to {Quantum} {Criticality}
  in {AFe$_2$As$_2$} with {A}={K}, {Rb}, and {Cs}}.
\newblock \emph{\bibinfo{journal}{Physical Review Letters}}
  \textbf{\bibinfo{volume}{116}}, \bibinfo{pages}{237003}
  (\bibinfo{year}{2016}).
\newblock
  \urlprefix\url{https://link.aps.org/doi/10.1103/PhysRevLett.116.237003}.

\bibitem{charnukha_intrinsic_2018}
\bibinfo{author}{Charnukha, A.} \emph{et~al.}
\newblock \bibinfo{title}{Intrinsic {Charge} {Dynamics} in {High}-{T$_c$}
  {AFeAs(O,F)} {Superconductors}}.
\newblock \emph{\bibinfo{journal}{Physical Review Letters}}
  \textbf{\bibinfo{volume}{120}}, \bibinfo{pages}{087001}
  (\bibinfo{year}{2018}).
\newblock
  \urlprefix\url{https://link.aps.org/doi/10.1103/PhysRevLett.120.087001}.

\bibitem{luo_electron_2013}
\bibinfo{author}{Luo, H.} \emph{et~al.}
\newblock \bibinfo{title}{Electron doping evolution of the magnetic excitations
  in {BaFe$_{2-x}$Ni$_x$As$_2$}}.
\newblock \emph{\bibinfo{journal}{Physical Review B}}
  \textbf{\bibinfo{volume}{88}}, \bibinfo{pages}{144516}
  (\bibinfo{year}{2013}).
\newblock \urlprefix\url{http://link.aps.org/doi/10.1103/PhysRevB.88.144516}.

\bibitem{dean_persistence_2013}
\bibinfo{author}{Dean, M. P.~M.} \emph{et~al.}
\newblock \bibinfo{title}{Persistence of magnetic excitations in
  {La$_{2-x}$Sr$_x$CuO$_4$} from the undoped insulator to the heavily overdoped
  non-superconducting metal}.
\newblock \emph{\bibinfo{journal}{Nature Materials}}
  \textbf{\bibinfo{volume}{12}}, \bibinfo{pages}{1019--1023}
  (\bibinfo{year}{2013}).
\newblock
  \urlprefix\url{http://www.nature.com/nmat/journal/v12/n11/abs/nmat3723.html}.

\bibitem{le_tacon_intense_2011}
\bibinfo{author}{Le~Tacon, M.} \emph{et~al.}
\newblock \bibinfo{title}{Intense paramagnon excitations in a large family of
  high-temperature superconductors}.
\newblock \emph{\bibinfo{journal}{Nature Physics}}
  \textbf{\bibinfo{volume}{7}}, \bibinfo{pages}{725--730}
  (\bibinfo{year}{2011}).
\newblock
  \urlprefix\url{http://www.nature.com/nphys/journal/v7/n9/full/nphys2041.html}.

\bibitem{lee_asymmetry_2014}
\bibinfo{author}{Lee, W.~S.} \emph{et~al.}
\newblock \bibinfo{title}{Asymmetry of collective excitations in electron- and
  hole-doped cuprate superconductors}.
\newblock \emph{\bibinfo{journal}{Nature Physics}}
  \textbf{\bibinfo{volume}{10}}, \bibinfo{pages}{883--889}
  (\bibinfo{year}{2014}).
\newblock
  \urlprefix\url{http://www.nature.com/nphys/journal/v10/n11/abs/nphys3117.html}.

\bibitem{ishii_high-energy_2014}
\bibinfo{author}{Ishii, K.} \emph{et~al.}
\newblock \bibinfo{title}{High-energy spin and charge excitations in
  electron-doped copper oxide superconductors}.
\newblock \emph{\bibinfo{journal}{Nature Communications}}
  \textbf{\bibinfo{volume}{5}} (\bibinfo{year}{2014}).
\newblock
  \urlprefix\url{http://www.nature.com/ncomms/2014/140425/ncomms4714/abs/ncomms4714.html}.

\bibitem{dellea_spin_2017}
\bibinfo{author}{Dellea, G.} \emph{et~al.}
\newblock \bibinfo{title}{Spin and charge excitations in artificial hole- and
  electron-doped infinite layer cuprate superconductors}.
\newblock \emph{\bibinfo{journal}{Physical Review B}}
  \textbf{\bibinfo{volume}{96}}, \bibinfo{pages}{115117}
  (\bibinfo{year}{2017}).
\newblock \urlprefix\url{https://link.aps.org/doi/10.1103/PhysRevB.96.115117}.

\bibitem{das_intermediate_2014}
\bibinfo{author}{Das, T.}, \bibinfo{author}{Markiewicz, R.~S.} \&
  \bibinfo{author}{Bansil, A.}
\newblock \bibinfo{title}{Intermediate coupling model of the cuprates}.
\newblock \emph{\bibinfo{journal}{Advances in Physics}}
  \textbf{\bibinfo{volume}{63}}, \bibinfo{pages}{151--266}
  (\bibinfo{year}{2014}).
\newblock \urlprefix\url{http://dx.doi.org/10.1080/00018732.2014.940227}.

\bibitem{monney_resonant_2016}
\bibinfo{author}{Monney, C.} \emph{et~al.}
\newblock \bibinfo{title}{Resonant inelastic x-ray scattering study of the spin
  and charge excitations in the overdoped superconductor
  {La$_{1.77}$Sr$_{0.23}$CuO$_4$}}.
\newblock \emph{\bibinfo{journal}{Physical Review B}}
  \textbf{\bibinfo{volume}{93}}, \bibinfo{pages}{075103}
  (\bibinfo{year}{2016}).
\newblock \urlprefix\url{http://link.aps.org/doi/10.1103/PhysRevB.93.075103}.

\bibitem{guarise_anisotropic_2014}
\bibinfo{author}{Guarise, M.} \emph{et~al.}
\newblock \bibinfo{title}{Anisotropic softening of magnetic excitations along
  the nodal direction in superconducting cuprates}.
\newblock \emph{\bibinfo{journal}{Nature Communications}}
  \textbf{\bibinfo{volume}{5}} (\bibinfo{year}{2014}).
\newblock
  \urlprefix\url{http://www.nature.com/ncomms/2014/141218/ncomms6760/abs/ncomms6760.html}.

\bibitem{dean_itinerant_2014}
\bibinfo{author}{Dean, M. P.~M.} \emph{et~al.}
\newblock \bibinfo{title}{Itinerant effects and enhanced magnetic interactions
  in {Bi}-based multilayer cuprates}.
\newblock \emph{\bibinfo{journal}{Physical Review B}}
  \textbf{\bibinfo{volume}{90}}, \bibinfo{pages}{220506}
  (\bibinfo{year}{2014}).
\newblock \urlprefix\url{http://link.aps.org/doi/10.1103/PhysRevB.90.220506}.

\bibitem{gretarsson_spin-state_2013}
\bibinfo{author}{Gretarsson, H.} \emph{et~al.}
\newblock \bibinfo{title}{Spin-{State} {Transition} in the {Fe} {Pnictides}}.
\newblock \emph{\bibinfo{journal}{Physical Review Letters}}
  \textbf{\bibinfo{volume}{110}}, \bibinfo{pages}{047003}
  (\bibinfo{year}{2013}).
\newblock
  \urlprefix\url{http://link.aps.org/doi/10.1103/PhysRevLett.110.047003}.

\bibitem{gretarsson_revealing_2011}
\bibinfo{author}{Gretarsson, H.} \emph{et~al.}
\newblock \bibinfo{title}{Revealing the dual nature of magnetism in iron
  pnictides and iron chalcogenides using x-ray emission spectroscopy}.
\newblock \emph{\bibinfo{journal}{Physical Review B}}
  \textbf{\bibinfo{volume}{84}}, \bibinfo{pages}{100509}
  (\bibinfo{year}{2011}).
\newblock \urlprefix\url{http://link.aps.org/doi/10.1103/PhysRevB.84.100509}.

\bibitem{glatzel_high_2005}
\bibinfo{author}{Glatzel, P.} \& \bibinfo{author}{Bergmann, U.}
\newblock \bibinfo{title}{High resolution 1s core hole {X}-ray spectroscopy in
  3d transition metal complexes—electronic and structural information}.
\newblock \emph{\bibinfo{journal}{Coordination Chemistry Reviews}}
  \textbf{\bibinfo{volume}{249}}, \bibinfo{pages}{65--95}
  (\bibinfo{year}{2005}).
\newblock
  \urlprefix\url{http://www.sciencedirect.com/science/article/pii/S0010854504001146}.

\bibitem{vanko_probing_2006}
\bibinfo{author}{Vanko, G.} \emph{et~al.}
\newblock \bibinfo{title}{Probing the 3d {Spin} {Momentum} with {X}-ray
  {Emission} {Spectroscopy}:  {The} {Case} of {Molecular}-{Spin}
  {Transitions}}.
\newblock \emph{\bibinfo{journal}{The Journal of Physical Chemistry B}}
  \textbf{\bibinfo{volume}{110}}, \bibinfo{pages}{11647--11653}
  (\bibinfo{year}{2006}).
\newblock \urlprefix\url{http://dx.doi.org/10.1021/jp0615961}.

\bibitem{yamamoto_origin_2016}
\bibinfo{author}{Yamamoto, Y.} \emph{et~al.}
\newblock \bibinfo{title}{Origin of {Pressure}-induced {Superconducting}
  {Phase} in {K$_x$Fe$_{2-y}$Se$_2$} studied by {Synchrotron} {X}-ray
  {Diffraction} and {Spectroscopy}}.
\newblock \emph{\bibinfo{journal}{Scientific Reports}}
  \textbf{\bibinfo{volume}{6}}, \bibinfo{pages}{30946} (\bibinfo{year}{2016}).
\newblock
  \urlprefix\url{http://www.nature.com/srep/2016/160808/srep30946/full/srep30946.html}.

\bibitem{lafuerza_evidence_2017}
\bibinfo{author}{Lafuerza, S.} \emph{et~al.}
\newblock \bibinfo{title}{Evidence of {Mott} physics in iron pnictides from
  x-ray spectroscopy}.
\newblock \emph{\bibinfo{journal}{Physical Review B}}
  \textbf{\bibinfo{volume}{96}}, \bibinfo{pages}{045133}
  (\bibinfo{year}{2017}).
\newblock \urlprefix\url{https://link.aps.org/doi/10.1103/PhysRevB.96.045133}.

\bibitem{pelliciari_magnetic_2017}
\bibinfo{author}{Pelliciari, J.} \emph{et~al.}
\newblock \bibinfo{title}{Magnetic moment evolution and spin freezing in doped
  {BaFe$_2$As$_2$}}.
\newblock \emph{\bibinfo{journal}{Scientific Reports}}
  \textbf{\bibinfo{volume}{7}}, \bibinfo{pages}{8003} (\bibinfo{year}{2017}).
\newblock \urlprefix\url{https://www.nature.com/articles/s41598-017-07286-6}.

\bibitem{vilmercati_itinerant_2012}
\bibinfo{author}{Vilmercati, P.} \emph{et~al.}
\newblock \bibinfo{title}{Itinerant electrons, local moments, and magnetic
  correlations in the pnictide superconductors {CeFeAsO$_{1-x}$F$_x$} and
  {Sr(Fe$_{1-x}$Co$_x$)$_2$As$_2$}}.
\newblock \emph{\bibinfo{journal}{Physical Review B}}
  \textbf{\bibinfo{volume}{85}}, \bibinfo{pages}{220503}
  (\bibinfo{year}{2012}).
\newblock \urlprefix\url{http://link.aps.org/doi/10.1103/PhysRevB.85.220503}.

\bibitem{hu_structural_2015}
\bibinfo{author}{Hu, D.} \emph{et~al.}
\newblock \bibinfo{title}{Structural and {Magnetic} {Phase} {Transitions} near
  {Optimal} {Superconductivity} in {BaFe$_2$(As$_{1-x}$P$_x$)$_2$}}.
\newblock \emph{\bibinfo{journal}{Physical Review Letters}}
  \textbf{\bibinfo{volume}{114}}, \bibinfo{pages}{157002}
  (\bibinfo{year}{2015}).
\newblock
  \urlprefix\url{http://link.aps.org/doi/10.1103/PhysRevLett.114.157002}.

\bibitem{allred_coincident_2014}
\bibinfo{author}{Allred, J.~M.} \emph{et~al.}
\newblock \bibinfo{title}{Coincident structural and magnetic order in
  {BaFe$_2$(As$_{1-x}$P$_x$)$_2$} revealed by high-resolution neutron
  diffraction}.
\newblock \emph{\bibinfo{journal}{Physical Review B}}
  \textbf{\bibinfo{volume}{90}}, \bibinfo{pages}{104513}
  (\bibinfo{year}{2014}).
\newblock \urlprefix\url{http://link.aps.org/doi/10.1103/PhysRevB.90.104513}.

\bibitem{kasahara_evolution_2010}
\bibinfo{author}{Kasahara, S.} \emph{et~al.}
\newblock \bibinfo{title}{Evolution from non-{Fermi}- to {Fermi}-liquid
  transport via isovalent doping in {BaFe$_2$(As$_{1-x}$P$_x$)$_2$}
  superconductors}.
\newblock \emph{\bibinfo{journal}{Physical Review B}}
  \textbf{\bibinfo{volume}{81}}, \bibinfo{pages}{184519}
  (\bibinfo{year}{2010}).
\newblock \urlprefix\url{http://link.aps.org/doi/10.1103/PhysRevB.81.184519}.

\bibitem{nakajima_growth_2012}
\bibinfo{author}{Nakajima, M.} \emph{et~al.}
\newblock \bibinfo{title}{Growth of {BaFe$_2$(As$_{1-x}$P$_x$)$_2$} {Single}
  {Crystals} (0$\leq$x$\leq$1) by {Ba$_2$As$_3$/Ba$_2$P$_3$}-{Flux} {Method}}.
\newblock \emph{\bibinfo{journal}{Journal of the Physical Society of Japan}}
  \textbf{\bibinfo{volume}{81}}, \bibinfo{pages}{104710}
  (\bibinfo{year}{2012}).
\newblock
  \urlprefix\url{http://journals.jps.jp/doi/abs/10.1143/JPSJ.81.104710}.

\bibitem{strocov_high-resolution_2010}
\bibinfo{author}{Strocov, V.~N.} \emph{et~al.}
\newblock \bibinfo{title}{High-resolution soft {X}-ray beamline {ADRESS} at the
  {Swiss} {Light} {Source} for resonant inelastic {X}-ray scattering and
  angle-resolved photoelectron spectroscopies}.
\newblock \emph{\bibinfo{journal}{Journal of Synchrotron Radiation}}
  \textbf{\bibinfo{volume}{17}}, \bibinfo{pages}{631--643}
  (\bibinfo{year}{2010}).
\newblock \urlprefix\url{http://www.ncbi.nlm.nih.gov/pmc/articles/PMC2927903/}.

\bibitem{ghiringhelli_saxes_2006}
\bibinfo{author}{Ghiringhelli, G.} \emph{et~al.}
\newblock \bibinfo{title}{{SAXES}, a high resolution spectrometer for resonant
  x-ray emission in the 400–1600 ev energy range}.
\newblock \emph{\bibinfo{journal}{Review of Scientific Instruments}}
  \textbf{\bibinfo{volume}{77}}, \bibinfo{pages}{113108}
  (\bibinfo{year}{2006}).
\newblock
  \urlprefix\url{http://scitation.aip.org/content/aip/journal/rsi/77/11/10.1063/1.2372731}.

\bibitem{hancock_evidence_2010}
\bibinfo{author}{Hancock, J.~N.} \emph{et~al.}
\newblock \bibinfo{title}{Evidence for core-hole-mediated inelastic x-ray
  scattering from metallic {Fe$_{1.087}$Te}}.
\newblock \emph{\bibinfo{journal}{Physical Review B}}
  \textbf{\bibinfo{volume}{82}}, \bibinfo{pages}{020513}
  (\bibinfo{year}{2010}).
\newblock \urlprefix\url{http://link.aps.org/doi/10.1103/PhysRevB.82.020513}.

\bibitem{graser_near-degeneracy_2009}
\bibinfo{author}{Graser, S.}, \bibinfo{author}{Maier, T.~A.},
  \bibinfo{author}{Hirschfeld, P.~J.} \& \bibinfo{author}{Scalapino, D.~J.}
\newblock \bibinfo{title}{Near-degeneracy of several pairing channels in
  multiorbital models for the {Fe} pnictides}.
\newblock \emph{\bibinfo{journal}{New Journal of Physics}}
  \textbf{\bibinfo{volume}{11}}, \bibinfo{pages}{025016}
  (\bibinfo{year}{2009}).
\newblock
  \urlprefix\url{http://stacks.iop.org/1367-2630/11/i=2/a=025016?key=crossref.a72831a9942b2bd8efb605fbd013abf2}.

\bibitem{richard_fe-based_2011}
\bibinfo{author}{Richard, P.}, \bibinfo{author}{Sato, T.},
  \bibinfo{author}{Nakayama, K.}, \bibinfo{author}{Takahashi, T.} \&
  \bibinfo{author}{Ding, H.}
\newblock \bibinfo{title}{Fe-based superconductors: an angle-resolved
  photoemission spectroscopy perspective}.
\newblock \emph{\bibinfo{journal}{Reports on Progress in Physics}}
  \textbf{\bibinfo{volume}{74}}, \bibinfo{pages}{124512}
  (\bibinfo{year}{2011}).
\newblock \urlprefix\url{http://stacks.iop.org/0034-4885/74/i=12/a=124512}.

\bibitem{yoshida_two-dimensional_2011}
\bibinfo{author}{Yoshida, T.} \emph{et~al.}
\newblock \bibinfo{title}{Two-{Dimensional} and {Three}-{Dimensional} {Fermi}
  {Surfaces} of {Superconducting} {BaFe$_2$(As$_{1-x}$P$_x$)$_2$} and {Their}
  {Nesting} {Properties} {Revealed} by {Angle}-{Resolved} {Photoemission}
  {Spectroscopy}}.
\newblock \emph{\bibinfo{journal}{Physical Review Letters}}
  \textbf{\bibinfo{volume}{106}}, \bibinfo{pages}{117001}
  (\bibinfo{year}{2011}).
\newblock
  \urlprefix\url{http://link.aps.org/doi/10.1103/PhysRevLett.106.117001}.

\bibitem{arnold_nesting_2011}
\bibinfo{author}{Arnold, B.~J.} \emph{et~al.}
\newblock \bibinfo{title}{Nesting of electron and hole {Fermi} surfaces in
  nonsuperconducting {BaFe}\$\{\}\_\{2\}\${P}\$\{\}\_\{2\}\$}.
\newblock \emph{\bibinfo{journal}{Physical Review B}}
  \textbf{\bibinfo{volume}{83}}, \bibinfo{pages}{220504}
  (\bibinfo{year}{2011}).
\newblock \urlprefix\url{http://link.aps.org/doi/10.1103/PhysRevB.83.220504}.

\bibitem{analytis_enhanced_2010}
\bibinfo{author}{Analytis, J.~G.}, \bibinfo{author}{Chu, J.-H.},
  \bibinfo{author}{McDonald, R.~D.}, \bibinfo{author}{Riggs, S.~C.} \&
  \bibinfo{author}{Fisher, I.~R.}
\newblock \bibinfo{title}{Enhanced {Fermi}-{Surface} {Nesting} in
  {Superconducting} {BaFe$_2$(As$_{1-x}$P$_x$)$_2$} {Revealed} by the de
  {Haas}{\textbackslash}char21\{\}van {Alphen} {Effect}}.
\newblock \emph{\bibinfo{journal}{Physical Review Letters}}
  \textbf{\bibinfo{volume}{105}}, \bibinfo{pages}{207004}
  (\bibinfo{year}{2010}).
\newblock
  \urlprefix\url{http://link.aps.org/doi/10.1103/PhysRevLett.105.207004}.

\bibitem{das_origin_2013}
\bibinfo{author}{Das, T.} \& \bibinfo{author}{Balatsky, A.~V.}
\newblock \bibinfo{title}{Origin of pressure induced second superconducting
  dome in {A$_y$Fe$_{2-x}$Se$_2$} [{A} = {K}, ({Tl},{Rb})]}.
\newblock \emph{\bibinfo{journal}{New Journal of Physics}}
  \textbf{\bibinfo{volume}{15}}, \bibinfo{pages}{093045}
  (\bibinfo{year}{2013}).
\newblock \urlprefix\url{http://stacks.iop.org/1367-2630/15/i=9/a=093045}.

\end{thebibliography}

\clearpage
\end{document}